\documentclass[prd,aps,preprint,nofootinbib,superscriptaddress]{revtex4}

\usepackage{amsmath}
\usepackage{epsfig}

\def\sv{$\langle\sigma v\rangle_0$}

\begin{document}

\title{The Quest for Supersymmetry: Early LHC Results versus\\ Direct and Indirect Neutralino Dark Matter Searches}

\author{Stefano Profumo}
\email{profumo@scipp.ucsc.edu} \affiliation{Department of Physics, University of California, 1156 High St., Santa Cruz, CA 95064, USA}\affiliation{Santa Cruz Institute for Particle Physics, Santa Cruz, CA 95064, USA}

\date{\today}

\begin{abstract}
\noindent We compare the first results on searches for supersymmetry with the Large Hadron Collider  (LHC) to the current and near-term performance of experiments sensitive to neutralino dark matter. We limit our study to the particular slices of parameter space of the constrained minimal supersymmetric extension to the Standard Model where CMS and ATLAS exclusion limits have been presented so far. We show where, on that parameter space, the lightest neutralino possesses a thermal relic abundance matching the value inferred by cosmological observations. We then calculate rates for, and estimate the performance of, experiments sensitive to direct and indirect signals from neutralino dark matter. We argue that this is a unique point in time, where the quest for supersymmetry -- at least in one of its practical and simple incarnations -- is undergoing a close scrutiny from the LHC and from dark matter searches that is both synergistic and complementary. Should the time of discovery finally unravel, the current performances of the collider program and of direct and indirect dark matter searches are at a conjuncture offering unique opportunities for a breakthrough on the nature of physics beyond the Standard Model.
\end{abstract}

\maketitle

\section{Introduction}
The present time is one ripe for discovery in the quest for physics beyond the Standard Model and for the nature of dark matter. A remarkable convergence of experimental searches for astrophysical and for laboratory signatures from dark matter is taking place at the same time when the Large Hadron Collider (LHC) is rapidly probing the parameter space of theories beyond the Standard Model associated to particle dark matter candidates. Of such theories, weak-scale supersymmetry \cite{susy} stands out as one of the most compelling frameworks.

The lack of knowledge of the mechanism of supersymmetry breaking implies the need for a generic parameterization of the weak-scale Lagrangian of the minimal supersymmetric extension of the Standard Model that contains more than a hundred parameters \cite{susyweakscale}. Imposing certain theory priors, such as the value of a handful of soft supersymmetry breaking parameters at the grand unification (GUT) scale, it is nonetheless possible to quantitatively compare the current performance of dark matter search experiments and of the collider program in the search for supersymmetry on a manageable yet representative parameter space. 

Quite remarkably, we shall show in this study that even in a very limited such slice of the supersymmetric parameter space, there exists a very high degree of {\em complementarity} between direct and indirect searches for dark matter and the current results on searches for supersymmetry with the ATLAS and CMS detectors at the LHC. Here, by ``complementarity'' we mean that relevant, viable parameter space regions exist that are excluded by each one of the mentioned search techniques but not by the others. Several parameter space regions also exist where more than one technique probes the associated particle physics setup. This {\em redundancy} of information and experimental techniques is as remarkable, and it will be as crucial as the complementarity mentioned above in unveiling the nature of particle physics beyond the Standard Model.

At present, searches for supersymmetry and for dark matter have not yielded yet clear discoveries, although several tantalizing experimental and observational reports have been produced in both indirect and direct dark matter searches. While further experimental data are needed to confirm or reject claims of discovery, conservatively it is at present at least possible to draw exclusion regions on the relevant theory parameter space. In a forthcoming time of discovery, however, the complementarity and redundancy of collider and dark matter searches for weak-scale supersymmetry might very well bring about unique opportunities to study and to disentangle properties of both the dark matter particle and the beyond-the-Standard-Model framework where such a particle lives.

In this study, we intend to compare the current performance of the dark matter search program with early LHC searches for supersymmetry. To this end, we start in the first part of the analysis by focusing on a very special corner of the weak-scale supersymmetry parameter space, namely the $\tan\beta=3$ slice of the so-called constrained minimal supersymmetric extension of the Standard Model (CMSSM), or minimal supergravity (mSUGRA), framework \cite{msugra_original, msugra_more, msugra}. This choice is solely motivated by the fact that this was the parameter space singled out and utilized by the ATLAS and CMS Collaborations to present the sensitivity reach of early searches for supersymmetry \cite{cmsjets,cmsdilepton,atlasj,atlasjpl,atlasbjets}, broadly based on an integrated luminosity of 35 ${\rm pb}^{-1}$ with a center of mass energy of 7 TeV.

Recent studies that have addressed the question of the impact of recent LHC results on dark matter direct and indirect detection experiments include Ref.~\cite{Akula:2011dd}, where the Authors studied what portion of the proton spin independent cross section versus the neutralino mass plane has been constrained by the recent LHC results, and further showed what part of the the relevant supersymmetric theory parameter space has been constrained by the recent XENON100 data \cite{xenon100}.  Ref.~\cite{Buchmueller:2011ki} and Ref.~\cite{Farina:2011bh} also addressed, in the context of a frequentist and of a best-fit analysis, respectively, a similar question.

The choice of very low $\tan\beta\lesssim 10$ tends to be in general highly unfavorable to dark matter searches (as well as also being in tension with LEP searches for the Higgs boson \cite{lephiggs}). For example, the relevant couplings for direct detection, and for the pair-annihilation cross section associated to some indirect detection channels, are generically suppressed in the low $\tan\beta$ regime. We do find, nevertheless, that current direct detection limits, specifically the recent XENON100 results \cite{xenon100}, even for $\tan\beta=3$ exhibit a significant degree of overlap with the performance of LHC searches for supersymmetry. Put in another way, the two experimental avenues are carving into the theory parameter space producing similar exclusion limits for this particular slice of theory parameter space. 

In exploring the phenomenology of dark matter on the parameter space regions illustratively employed by the LHC Collaborations, we also found an interesting and rather unique feature associated to a very important indirect detection channel, the direct pair-annihilation of dark matter into two monochromatic photons. The standard lore is that this annihilation channel is typically suppressed by the square of the electromagnetic fine-structure constant, giving a branching ratio into two photons not larger than $10^{-3}$. In the low $\tan\beta$ regime, however, we find that for lightest neutralino masses below the top-quark mass, other annihilation channels are so suppressed that the two-photon mode can be as large or larger than 0.3!

In the second part of this study, we explore two more slices of the CMSSM parameter space, at a larger value of $\tan\beta=45$, and for both signs of the $\mu$ Higgsino mass parameter, where dark matter searches (especially indirect searches) are more promising. We utilize here the results of Ref.~\cite{nath}, that extrapolated the LHC reach for $\tan\beta=45$ from the $\tan\beta=3$ results, and the results from the ATLAS search for $b$-jets and missing transverse momentum final states, relevant at large $\tan\beta$ \cite{atlasbjets}. In the large $\tan\beta$ regime, we exemplify a quite remarkable complementarity and redundancy in the regions being explored by the collider program with the LHC and by dark matter search experiments.

Even within the very limited and constrained theoretical setup of the CMSSM,  the present study draws three main conclusions:
\begin{enumerate}
\item regions of parameter space exist where only one technique is currently sensitive to supersymmetry and to supersymmetric dark matter, 
\item current and future experimental capabilities will lead to several regions of redundant overlap across detection techniques for supersymmetry and supersymmetric dark matter, that will be crucial in the time of discovery to pinpoint and understand the particle properties of dark matter and of the beyond the standard Model framework where such particle lives, and
\item in the event of a discovery at the LHC that points towards a BSM setup similar to the low-energy supersymmetric scenario under consideration here, the complementarity and partial overlap of different dark matter search techniques will be key to selecting the correct region of parameter space or to rule out alternatives.
\end{enumerate}

The ensuing study is organized as follows: in the next section we focus on the $\tan\beta=3$ case, and we explore in detail the properties of the lightest supersymmetric particle as a dark matter candidate, including the associated dark matter detection rates. Sec.~\ref{sec:45} compares the performance of collider, direct and indirect detection searches for supersymmetry and supersymmetric dark matter for $
\tan\beta=45$. The final Sec.~\ref{sec:disc} presents our discussion and conclusions.

\section{The CMSSM parameter space at $\tan\beta=3$}

In a series of recent papers \cite{cmsjets,cmsdilepton,atlasj,atlasjpl,atlasbjets}, the CMS and ATLAS collaborations presented the results of searches for supersymmetry with the first $\approx35\ {\rm pb}^{-1}$ of LHC data in $\sqrt{s}=7$ TeV proton-proton collisions. While none of the searches was specifically optimized for a particular supersymmetric model, the results were presented in all papers for the constrained minimal supersymmetric extension of the Standard Model (CMSSM, sometimes also referred to as minimal supergravity (mSUGRA) model; the latter name, however, is also used by some investigators to indicate an even more constrained scenario \cite{msugravscmssm}; for brevity, we will hereafter use only CMSSM instead of the lengthier acronym CMSSM/mSUGRA). The CMSSM model is parameterized by the values of three quantities defined at the GUT scale (the universal scalar mass $m_0$, the universal gaugino mass $m_{1/2}$ and the universal trilinear scalar coupling $A_0$) plus two parameters specified at the electro-weal scale: the ratio of the vacuum expectation values of the two neutral components of the two Higgs doublets, $\tan\beta$, and the sign of the Higgs mixing parameter $\mu$.

For the sake of comparing with the parameter space slice selected by the CMS \cite{cmsjets,cmsdilepton} and ATLAS \cite{atlasj,atlasjpl} collaborations to present the results of their early searches for supersymmetry, we focus on the $\tan\beta=3$, $A_0=0$ and $\mu>0$ slice of parameter space of the CMSSM \cite{msugra}. We employ, here and in the remainder of this study, the ISAJET package (v.7.78) \cite{isajet} to calculate the renormalization group evolution of parameters from the grand unification scale down to the electro-weak scale, as well as the particle spectrum. We use the DarkSUSY package (v.5.0.5) \cite{ds} to calculate dark matter detection rates. We refer the Reader to the DarkSUSY manual for details on assumptions specific to both direct and indirect detection rates -- unless otherwise specified, the choices we make here correspond to the default choices in DarkSUSY \cite{ds}.

We warn the Reader that different choices of assumptions and parameters -- including, but not limited to, the dark matter density profile and velocity distribution of the Milky Way and of the relevant dwarf galaxies, the strange-quark content of the proton, nuclear form factors, neutrino telescope maximum half-aperture angle -- would affect {\em significantly} the predictions we make of the current experimental sensitivities and the resulting reach in the theory parameter space. A detailed analysis of all of the relevant uncertainties is, however, beyond the scope of this work.

\begin{figure*}[!t]
\mbox{\includegraphics[width=0.5\textwidth,clip]{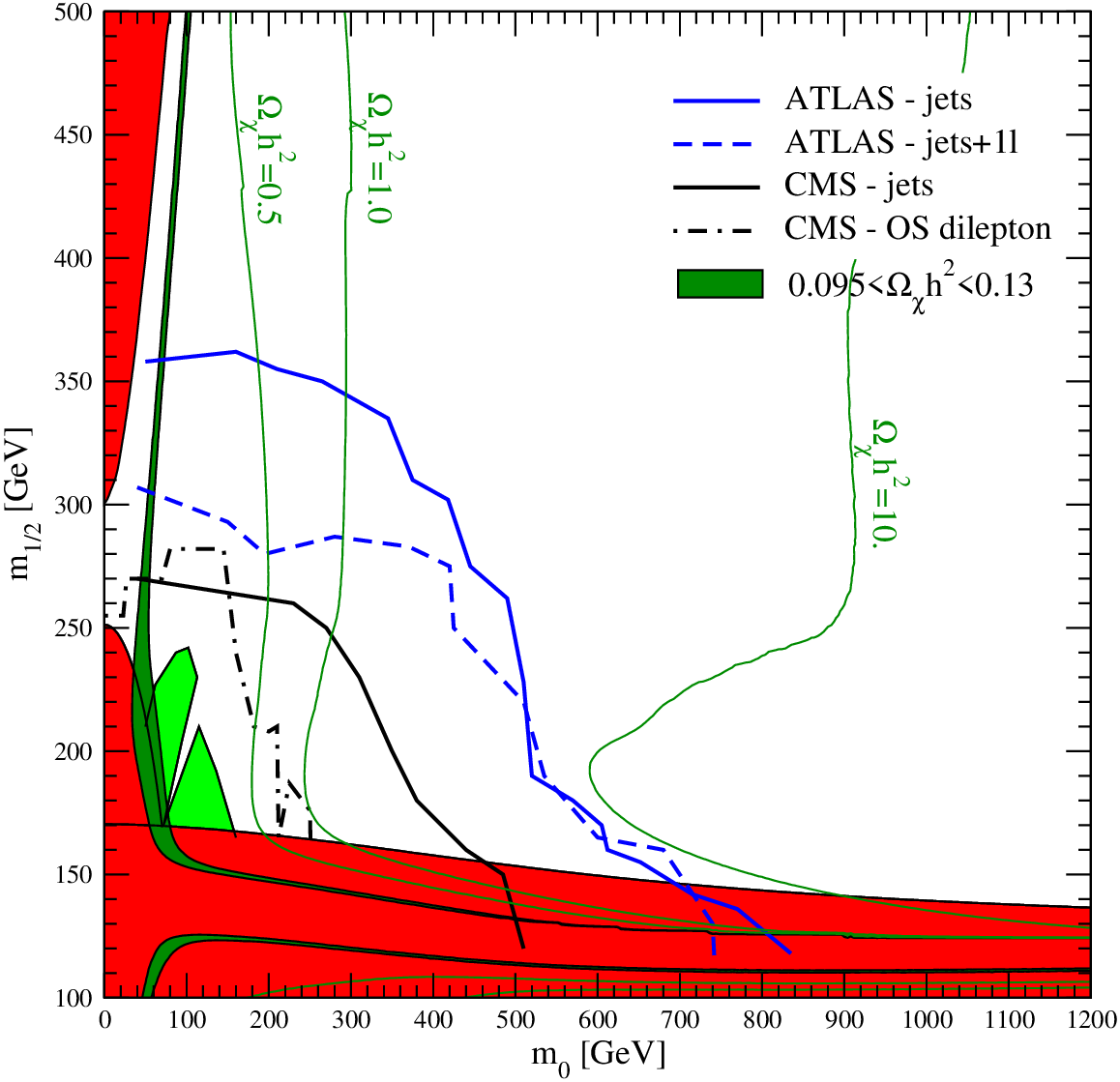}\quad\quad\includegraphics[width=0.5\textwidth,clip]{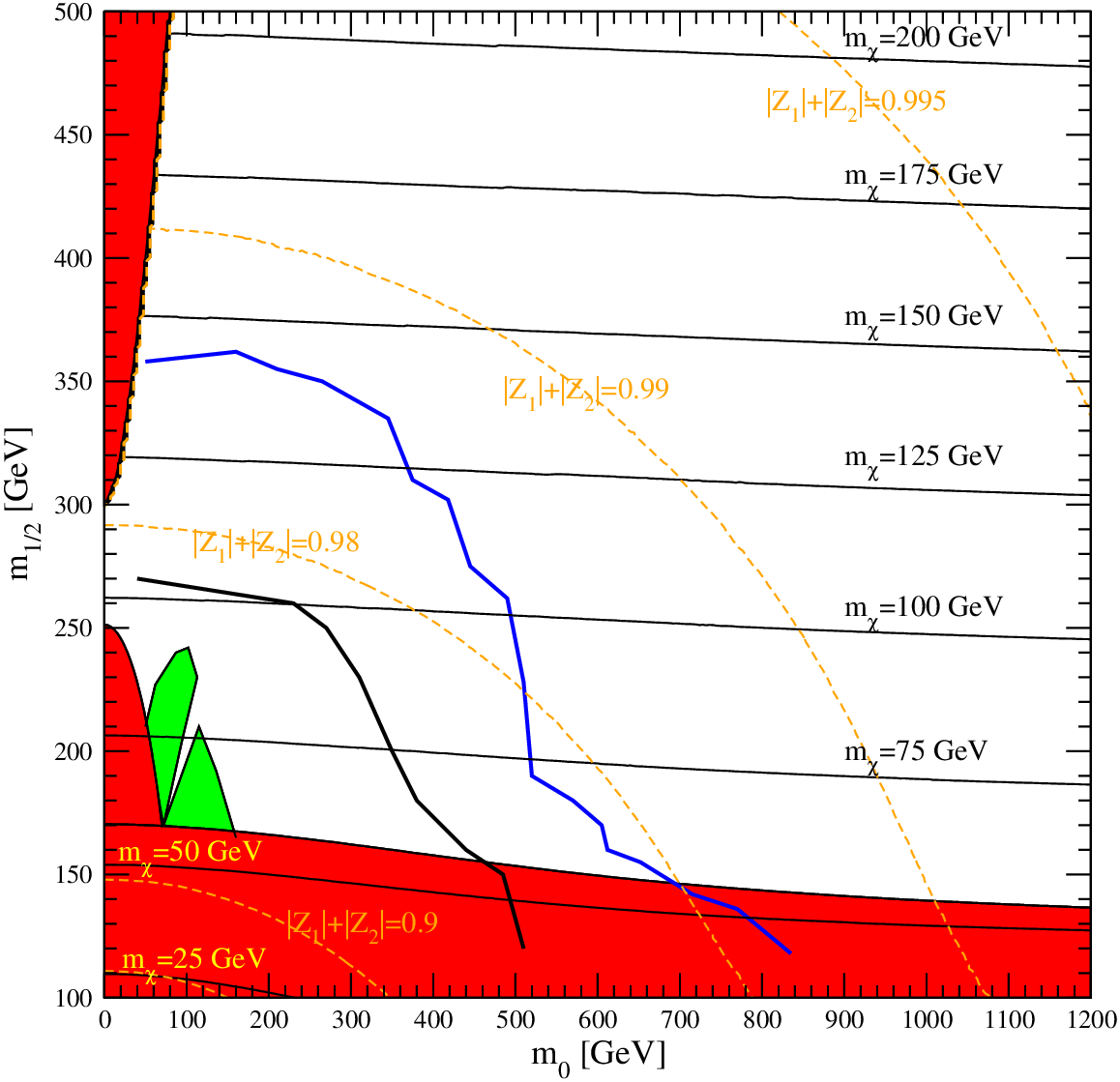}}
\caption{\label{fig:neut_tb03}\it\small Left: Curves at constant neutralino relic abundance $\Omega_\chi h^2$, on the $(m_0,m_{1/2})$ plane at $\mu>0$, $A_0=0$ and $\tan\beta=3$. The dark green regions have a relic neutralino density in the range $0.095<\Omega_\chi h^2<0.13$. The red regions are excluded by searches at LEP-II for charginos \cite{lep2charginos} (bottom), and for staus \cite{lep2staus} (bottom-left), or have a stau LSP (upper-left). The blue lines indicate the early ATLAS exclusion curves on this plane \cite{atlasj, atlasjpl}, while the black lines those from CMS \cite{cmsjets, cmsdilepton}. Right: On the same plane, curves of constant neutralino mass $m_\chi$ and constant bino plus wino fraction $|Z_1|+|Z_2|$ are shown.}
\end{figure*}

We reproduce in Fig.~\ref{fig:neut_tb03} the curves corresponding to the 95\% CL limits obtained searching for squarks and gluinos with final states containing jets plus missing transverse momentum at ATLAS \cite{atlasj} (solid blue line), searching for one-lepton, jets and missing transverse momentum with ATLAS \cite{atlasjpl} (dashed blue line), jets and missing transverse energy with the CMS detector \cite{cmsjets} (black solid line), and finally opposite-sign dilepton events with CMS \cite{cmsdilepton} (dot-dashed black line).

In Fig.~\ref{fig:neut_tb03}, we also shade in bright green the region excluded by the Tevatron D0 experiment in searches for associated chargino-neutralino production in trilepton final states \cite{dzero}. The red regions reproduce other constraints on the parameter space shown, namely: (1) in the top left part of the plot the lightest stau is the (stable) lightest supersymmetric particle (this is excluded by searches for a stable massive charged relic \cite{chargedrelic}); (2) in the lower left corner the lightest stau is lighter than the lower limit placed by searches for stau pair-production at LEP2 \cite{lep2staus} (3) in the lower part of the plot the chargino is lighter than the LEP2 lower limit \cite{lep2charginos}. 

We stress that, as in Ref.~\cite{cmsjets,cmsdilepton,atlasj,atlasjpl}, we do not to show constraints from Higgs searches at LEP, from B-physics, such as those from supersymmetric contributions to the branching ratio $b\to s\gamma$ or $B_s\to\mu^+\mu^-$, or from the muon anomalous magnetic moment. The reason for this choice is that we are not focusing on putting comprehensive constraints on the specific theoretical framework of the CMSSM, as done in other studies (see for example Ref.~\cite{nath}); rather, we exclusively intend to compare the performance of direct searches for supersymmetric particles at the LHC with the performance of dark matter experiments, and we utilize the CMSSM as a convenient theoretical scheme to carry out such a comparison. It is therefore beyond the scope of this analysis to compare LHC searches or dark matter search results with constraints from B-physics or from $(g-2)_\mu$.

The dark-green shaded contours show the regions approximately corresponding to a lightest neutralino thermal relic abundance matching the inferred abundance of cold dark matter from cosmological observations as analyzed including 7 years of WMAP data \cite{wmap7yrs} within a generous 5$\sigma$ range, i.e. $0.095<\Omega_\chi h^2<0.13$. Finally, in the left panel of Fig.~\ref{fig:neut_tb03} we indicate three curves corresponding to $\Omega_\chi h^2=0.5$, 1.0 and 10. 

The requirement that neutralinos have a thermal relic density matching the universal dark matter density should be regarded as a ``soft'' constraint: models with an under-abundant thermal density can easily accommodate having only neutralino dark matter if non-thermal dark matter production is considered -- this is not only possible, but very well theroertically motivated in a number of beyond-the-Standard Model frameworks (e.g. \cite{moroi}). Similarly, entropy injection can make regions with over-abundant thermal production also compatible with the observed dark matter density (see e.g. \cite{wainprofu}).

The thermal neutralino relic abundance is in accordance with the observed cold dark matter density in standard, well-known regions on the CMSSM parameter space (see e.g. \cite{ellisbaerpapers}): the vertical strip to the left corresponds to coannihilations with the lightest stau, while the two horizontal branches in the lower part of the plot (excluded by LEP2 chargino searches) correspond to rapid $s$-channel annihilation via the Standard-Model-like lightest neutral Higgs $h$, for $2m_\chi\sim m_h$. Structure in the iso-level curves corresponding  to heavier dark matter relic densities is associated to various thresholds for available final states for neutralino pair-annihilation, including electro-weak massive gauge bosons and the top quark.

In the right panel, we limit the LHC exclusion curves to the two best performing searches (ATLAS-jets and CMS-jets, blue and black solid curves, respectively), and indicate curves at constant lightest neutralino mass $m_\chi$ --- a quantity of great relevance for the ensuing phenomenology of the resulting dark matter candidate. On the parameter space shown, to a good degree of approximation, $m_\chi\simeq 0.4 m_{1/2}$, as expected from the renormalization group running of the hypercharge gaugino soft breaking mass $M_1$ when $\mu\gtrsim M_1$ and hence $m_\chi\simeq M_1$. 

In the right panel of Fig.~\ref{fig:neut_tb03}, we also show curves at constant $|Z_1|+|Z_2|$, where $Z_i$ indicates the neutralino mixing matrix element for the interaction eigenstates $i=\widetilde B,\ \widetilde W,\ \widetilde H_1,\ \widetilde H_2$. The quantity $|Z_1|+|Z_2|$ therefore indicates a measure of how ``gaugino-like'' the lightest neutralino is. As is well known, increasing the sfermion masses drives, via renormalization group evolution and the requirement of successful radiative electro-weak symmetry breaking, the $\mu$ parameter to smaller values, hence decreasing the gaugino character of the lightest, bino-like neutralino. Notice that the effect is rather small (at most at the percent level) for the small value of $\tan\beta=3$ under consideration here.

\begin{figure*}[!t]
\mbox{\includegraphics[width=0.5\textwidth,clip]{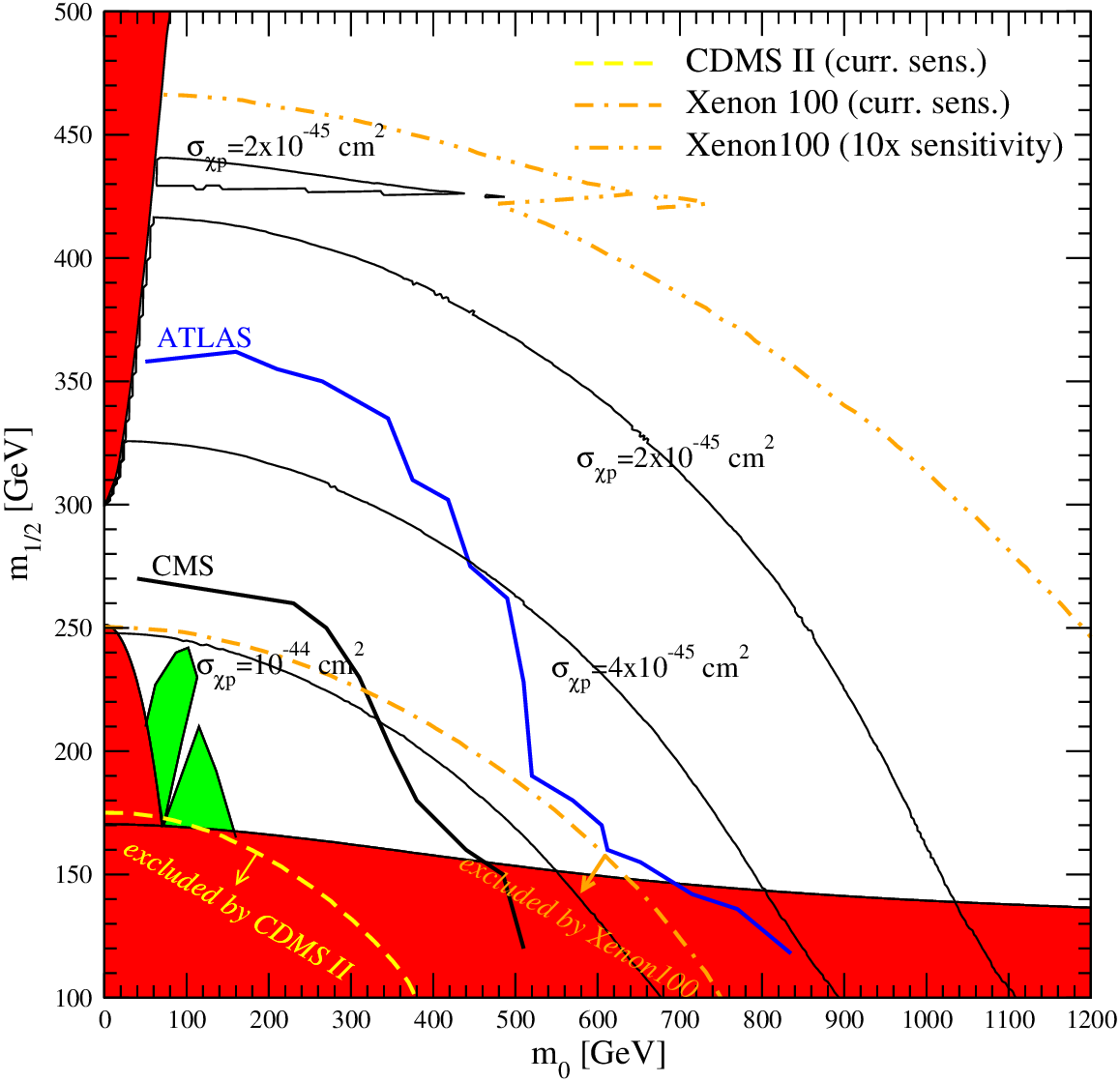}\quad\quad\includegraphics[width=0.5\textwidth,clip]{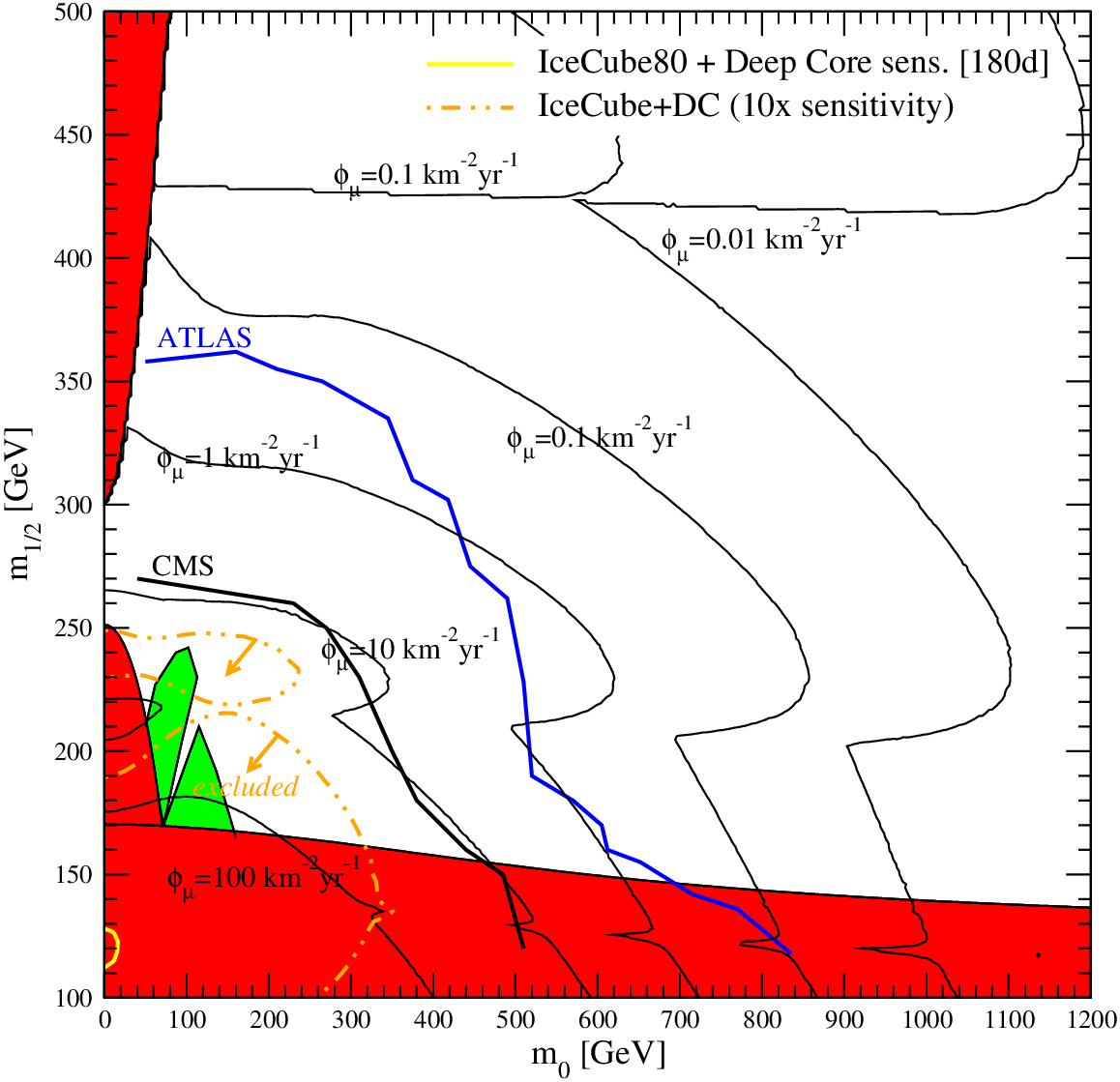}}
\caption{\label{fig:dirdet_tb03}\it\small Left: on the same parameter space as in Fig.~\ref{fig:neut_tb03}, black lines correspond to curves at constant spin-independent neutralino-proton scattering cross section. We also show the regions excluded by searches for dark matter with CDMS-II \cite{cdms2} (to the bottom left of the dashed yellow line) and with XENON100 \cite{xenon100} (dot-double-dashed orange line). Finally, we indicate the parameter space that would be excluded by a negative result with a sensitivity 10 times better than the current XENON100 results. Right: curves at constant flux of high-energy ($>1$ GeV) neutrinos from the Sun, and the anticipated sensitivity of 180 days of IceCube80 and Deep-Core data \cite{ icecube} (yellow contour, and 10 times the sensitivity, double-dot-dashed line).}
\end{figure*}

We begin our comparison of the performance of early LHC searches for supersymmetry with direct and indirect neutralino dark matter searches in Fig.~\ref{fig:dirdet_tb03}. The left panel (with the same red and light green excluded regions as in the previous region) shows, with black solid lines, points at constant values for the scalar neutralino-proton scattering cross section $\sigma_{\chi P}$. We also indicate the regions excluded by the most recent results from the CDMS II Collaboration \cite{cdms2} (dashed yellow line) and from the XENON100 Collaboration \cite{xenon100} (dot-double-dashed orange line). Notice that these exclusion region do not correspond to fixed values of the cross section $\sigma_{\chi P}$, since the regions also depend on the neutralino mass. We determine the curves by comparing, for each point in the parameter space, the predicted model spin-independent cross section with the experimentally determined limiting value corresponding to the calculated neutralino mass. The ``exclusion curve'' then corresponds to parameter space points where that ratio equals 1.

At low values of $\tan\beta$, and for the corresponding low higgsino fraction, the lightest neutralino coupling to the Higgs sector is very suppressed, leaving small contributions from squark exchange. Nonetheless, the recent XENON100 results represent a jump on this slice of the CMSSM parameter space from the CDMS II sensitivity comparable to the jump between the Tevatron and the early LHC. XENON100 excludes a comparable region of parameter space to the CMS results for relatively heavy neutralinos (large $m_{1/2}$), and to ATLAS for lighter neutralinos and heavier scalars. We also show, with a double-dotted-dashed orange line the exclusion limits that would correspond to an improvement in the XENON100 sensitivity by a factor 10. Such an improvement in sensitivity might be optimistic for XENON100 by the end of 2012, but is well within the anticipated sensitivity of the XENON1T experiment, recently approved by INFN to start at the Laboratori Nazionali del Gran Sasso \cite{elenaquote}, even with a very limited time exposure.

The right panel illustrates the expected flux of high-energy ($>1$ GeV) neutrinos from the Sun (we employ here a 1 GeV threshold to allow an easier comparison to the performance plots produced by the IceCube collaboration, see e.g. \cite{icecube}). The black solid lines correspond to curves at constant flux of neutrinos (several relevant pair-annihilation threshold are clearly visible, the most dramatic effect being associated to the $W$ boson and to the $t$ quark). We also consider the anticipated sensitivity for 180-days of data-taking for IceCube80 plus DeepCore \cite{icecube}, which, on this slice of the CMSSM parameter space, is limited to a tiny sliver at vanishing scalar masses and $m_{1/2}\sim120$ GeV, bounded by a yellow line. A jump by a factor ten in sensitivity would bring the excluded region to the double-dotted-dashed orange line, which is still far below the CMS exclusion region. We thus conclude that, unlike spin-independent direct detection, neutrino telescopes at low $\tan\beta$ are ineffective at cross-constraining collider searches for supersymmetry. 

\begin{figure*}[!t]
\mbox{\includegraphics[width=0.5\textwidth,clip]{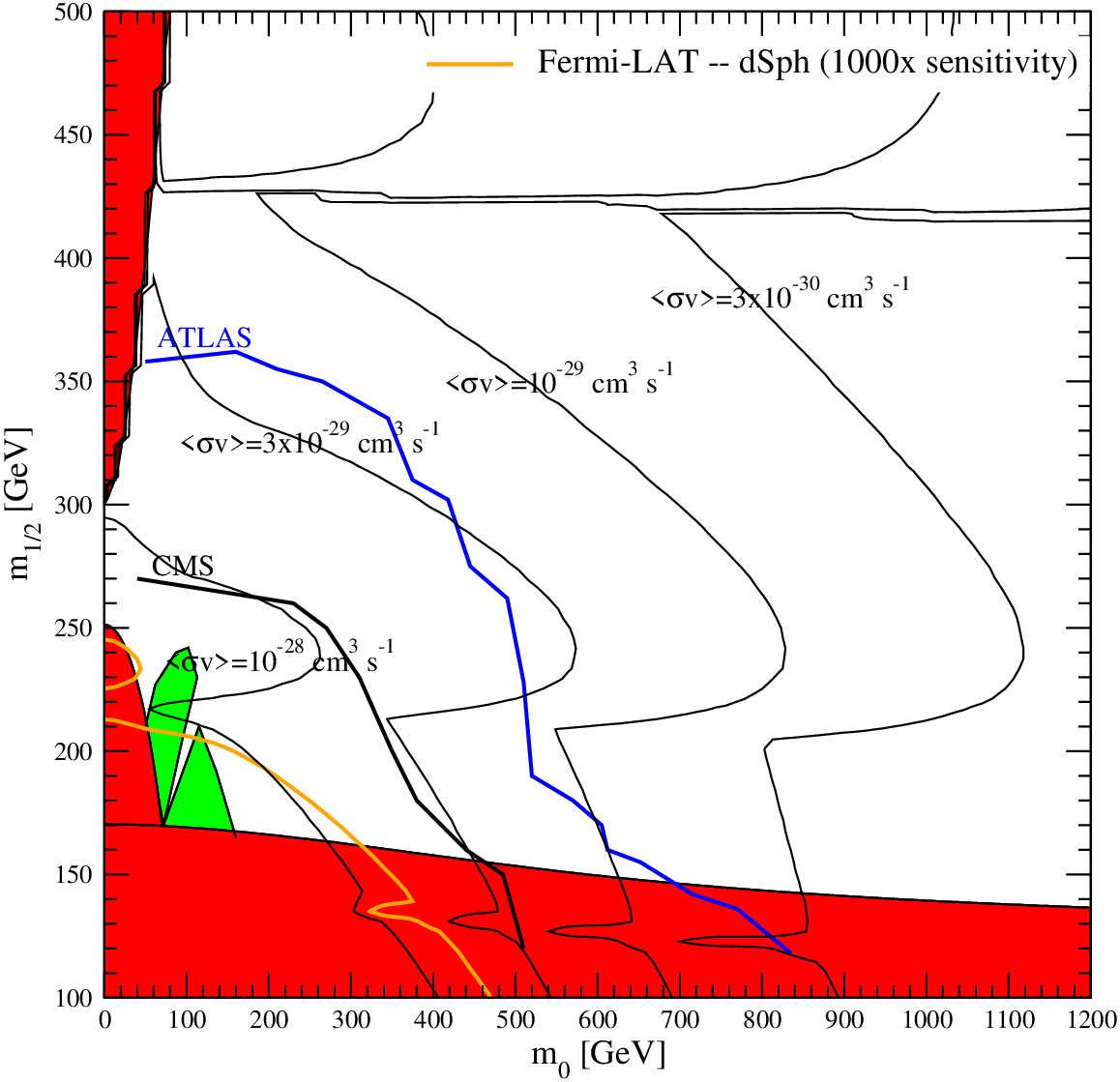}\quad\quad\includegraphics[width=0.5\textwidth,clip]{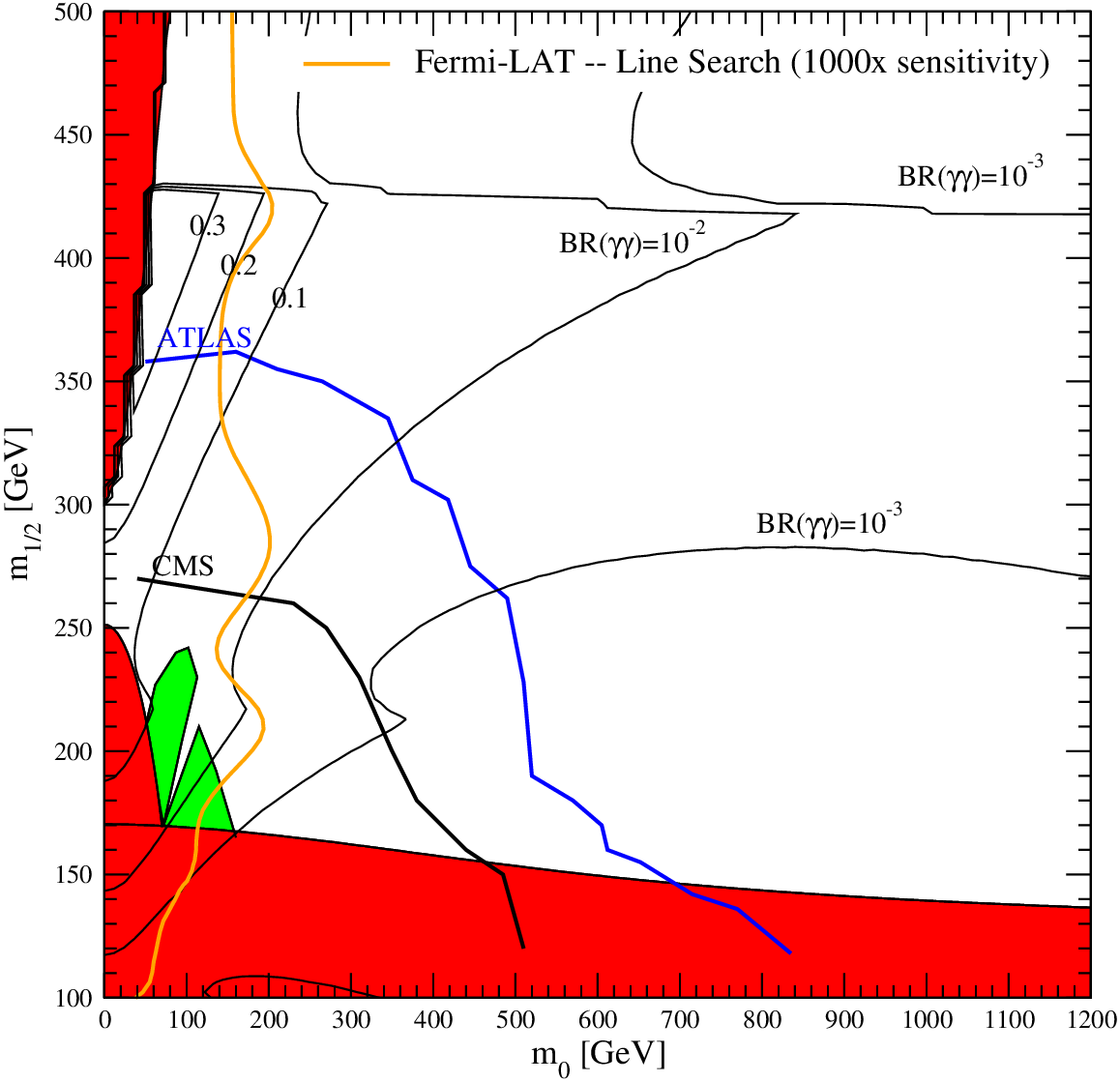}}
\caption{\label{fig:gammas_tb03}\it\small Left: On the same parameter space as in Fig.~\ref{fig:neut_tb03}, black lines correspond to curves at constant neutralino pair-annihilation cross section \sv. The orange line indicates the region corresponding to a sensitivity $10^3$ times larger than the current constraints from Fermi-LAT observations of nearby dSph \cite{dsph}. Right: curves at constant neutralino pair-annihilation branching ratio into two photons. The orange line indicates the region corresponding to a sensitivity $10^3$ times larger than the current constraints on the annihilation rate of dark matter into monochromatic photons \cite{grline}.}
\end{figure*}

A further promising indirect dark matter detection channel is the search for high-energy gamma rays from dark matter pair-annihilation. Here, we consider the results of searches for a gamma-ray signal from nearby dwarf spheroidal galaxies (dSph) with the Fermi Large Area Telescope (LAT), as reported in Ref.~\cite{dsph}. We point the Reader to Ref.~\cite{dsph} for details on the choice of targets, modeling of dark matter density profiles and related uncertainties, and background considerations. The performance of this detection technique for dark matter models depends on the dark matter pair-annihilation final state, and specifically on the rate of pair-annihilation and on the gamma-ray produced by given Standard Model final states. The amplitude for neutralino annihilation modes at zero-velocity into Standard Model fermions is proportional to the fermion mass, and the resulting annihilation cross section is parametrically suppressed by $m_f^2/m_\chi^2$. In the MSSM, the amplitudes for annihilation to down-type quarks and leptons, however, inherit a direct proportionality to $\tan\beta$ from their Yukawa couplings. As a result, in the low-$\tan\beta$ regime, and below the top-quark mass threshold (i.e. for $m_\chi<m_t$) neutralino pair-annihilation is highly suppressed, unless annihilation into gauge bosons is open and significant. This is not the typical case within the CMSSM, where the lightest neutralino is dominantly the fermionic partner to the U(1) hypercharge gauge boson, and therefore to a first approximation cannot pair-annihilate to $ZZ$ or $W^+W^-$ (unless it features a large higgsino fraction).

We show in Fig.~\ref{fig:gammas_tb03}, left, curves of constant values of the pair annihilation cross section times relative velocity, in the zero-velocity limit (a quantity we indicate with \sv). The Reader will notice that the values of the pair-annihilation cross section are well below what would be needed to produce a thermal relic density $\Omega_\chi h^2$ in accord with the observed cold dark matter density ($\langle\sigma v\rangle_0\sim3\times10^{-26}\ {\rm cm}^3 {\rm s}^{-1}$), since, here, $10^{-30}\lesssim\langle\sigma v\rangle_0/({\rm cm}^3 {\rm s}^{-1})\lesssim 10^{-28}$. This is indeed reflected in the values of $\Omega_\chi h^2$ we showed in Fig.~\ref{fig:neut_tb03} outside of the regions of parameter space where either stau coannihilations or resonant annihilation via the lightest Higgs $h$ are active. Notice that neither of these mechanisms leads to an enhancement in \sv, since staus have long decayed away at $T=0$, and the $h$ has the wrong CP assignment to pair-annihilate two Majorana fermions in an $s$ wave. The extremely low values of the pair annihilation cross section yield a very low gamma-ray from annihilations in astrophysical objects, such as dSph. Even low values of the lightest netrualino mass do not offset in the gamma-ray yield (scaling as $1/m_\chi^2$) the smallness of the annihilation rate. We find that the Fermi-LAT results on the non-observation of dSph in gamma rays translate into limits on \sv\ which are roughly a factor $10^3$ larger than the cross sections we find on the $\tan\beta=3$ parameter space under consideration. We illustrate this by showing a contour (orange solid line in the lower-right corner) corresponding to a limit $10^3$ times better than what found and reported in Ref.~\cite{dsph}.

A smoking-gun astrophysical signal of neutralino dark matter pair-annihilation is the monochromatic gamma-ray pair production resulting from prompt $\chi\chi\to\gamma\gamma$ processes. The Fermi-LAT Collaboration targeted in Ref.~\cite{grline} monochromatic gamma-ray lines analyzing the gamma-ray sky in suitable energy and angular regions, producing constraints on the maximal pair annihilation rate of dark matter into two photons, as a function of the dark matter mass. We use those limits (specifically with a Navarro-Frenk-White dark matter density profile \cite{nfw}) to put constraints on the CMSSM parameter space slice under consideration here. We show contours at constant values of the branching ratio for neutralino pair-annihilation into two photons in Fig.~\ref{fig:gammas_tb03}, right.

The $\chi\chi\to\gamma\gamma$ mode is loop-suppressed in supersymmetry, as neutralinos are neutral under U(1)${}_{\rm e.m.}$, so, in principle, the corresponding pair-annihilation branching ratio should be a factor ${\mathcal O}(\alpha_{\rm e.m.}^2/(4\pi))$ smaller than for other annihilation modes, with $\alpha_{\rm e.m.}$ the electromagnetic fine-structure constant. This naive estimate, however, neglects other suppression mechanisms that might apply to other annihilation final states, such as chirality suppression for light fermionic final states that we mentioned above. As a result, we find that the branching ratio into two photons is here exceptionally large, being on the order of 10\% or more along the co-annihilation strip (but below the top-quark threshold). The absolute value of the pair-annihilation cross section are however too small even for these large branching ratios. Again, we find that the line searches in the Fermi data fall short by roughly a factor $10^3$, as shown by the orange line corresponding to a sensitivity greater than the actual one by that factor. As expected, the best reach is where the branching ratio into two photons is large, i.e. along the coannihilation strip at low $m_0$.

\section{Large $\tan\beta$ and the complementarity of Direct and Indirect Dark Matter searches with the Collider program}\label{sec:45}

\begin{figure*}[!t]
\includegraphics[width=0.8\textwidth,clip]{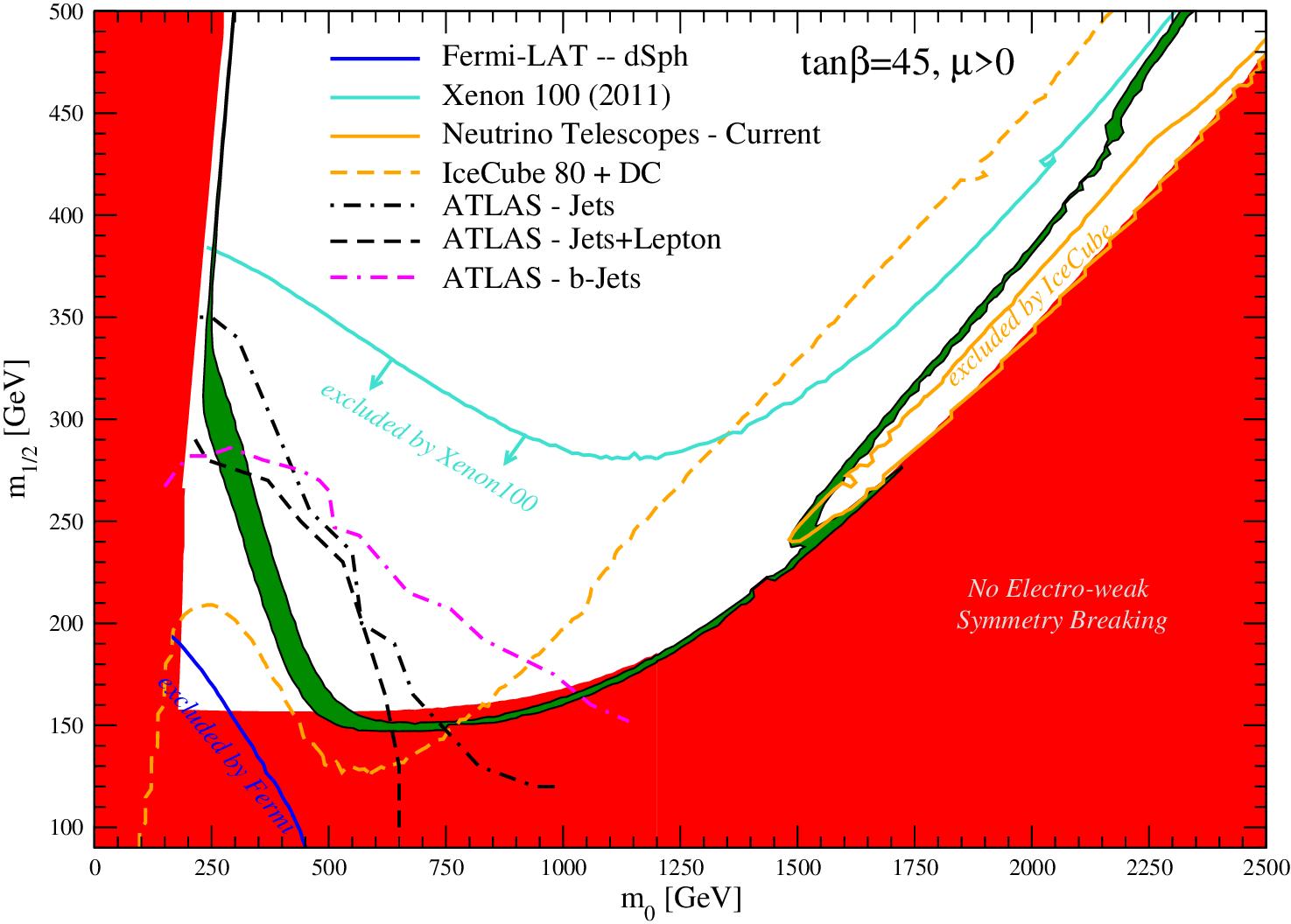}
\caption{\label{fig:tb45}\it\small The CMSSM $(m_0,m_{1/2})$ parameter space slice at $\tan\beta=45$ and $\mu>0$. The red region is ruled out by LEP-II searches \cite{lep2charginos, lep2staus}, or by the lack of successful radiative electro-weak symmetry breaking, or by featuring a charged LSP. The dark green region possesses  a relic neutralino density in the range $0.095<\Omega_\chi h^2<0.13$. The arrows on the exclusion lines points towards the {\em excluded} parameter space regions. We also include the $\tan\beta=40$ ATLAS b-jets combined exclusion region (magenta double-dashed-dotted line) from Ref.~\cite{atlasbjets}} 
\end{figure*}

\begin{figure*}[!t]
\includegraphics[width=0.8\textwidth,clip]{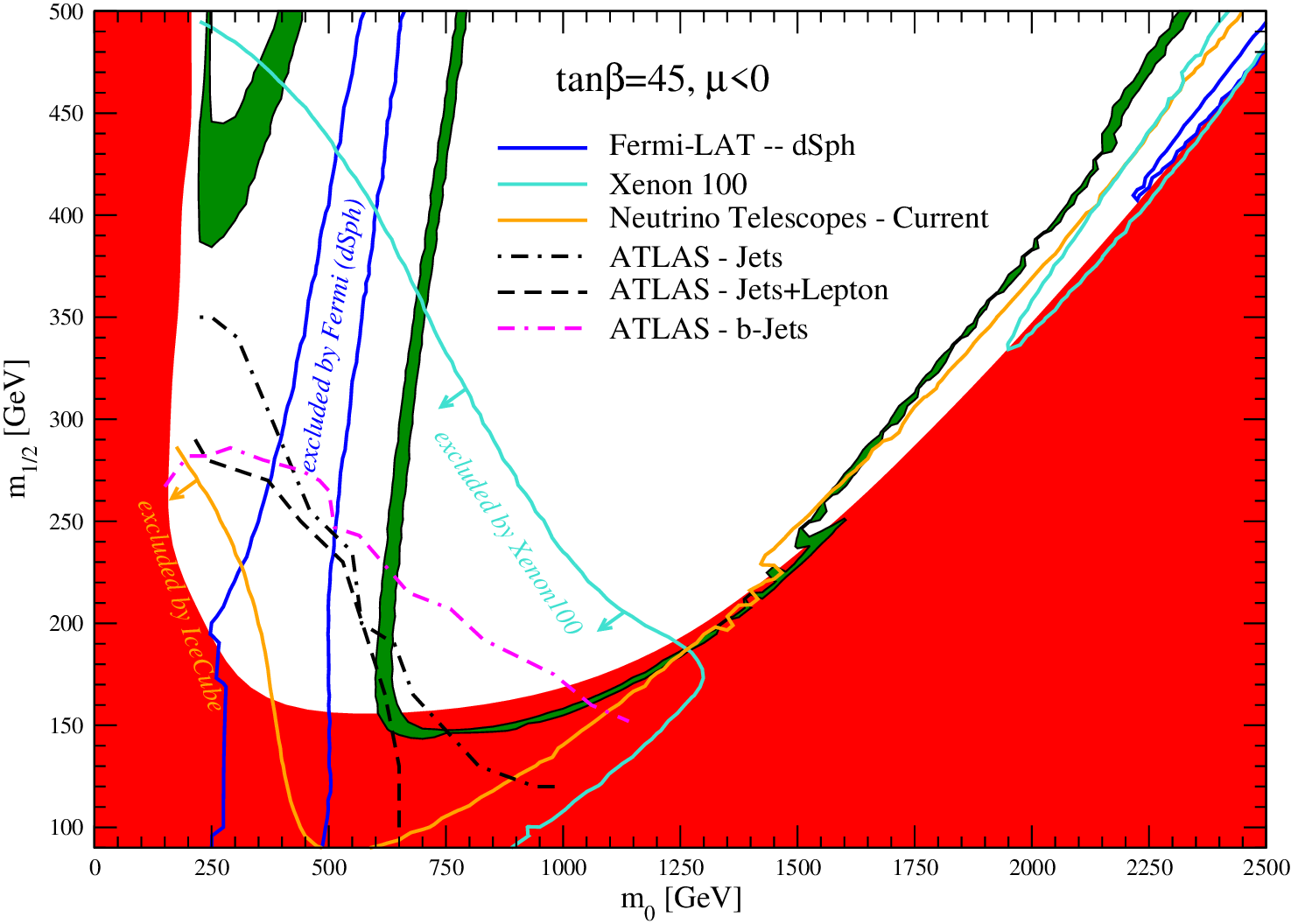}
\caption{\label{fig:tb45_muneg}\it\small As in Fig.~\ref{fig:tb45}, but for negative $\mu$.}
\end{figure*}

The phenomenology of the lightest neutralino as a dark matter particle depends quite steeply on the value of $\tan\beta$. For moderate to large values of $\tan\beta$, pair-annihilation into SU(2)${}_L$ down-type fermions is parametrically enhanced like $(\tan\beta)^2$, with larger rates for indirect detection; in addition, at large $\tan\beta$, due to the radiative electro-weak symmetry breaking conditions, the lightest neutralino typically has a larger higgsino fraction, with enhanced couplings to the Higgs sector; this impacts both direct and indirect detection rates. To illustrate how the current sensitivity of LHC experiments compares with the current performance of direct and indirect dark matter searches, we employ $\tan\beta=45$, and show the current exclusion regions for both positive and negative values of the $\mu$ parameter in Fig.~\ref{fig:tb45} and \ref{fig:tb45_muneg}, respectively.

For large $\tan\beta$, the ATLAS search for events with large missing transverse momentum and at least one $b$-jet \cite{atlasbjets} becomes very relevant, even out-performing other search channels, especially for large $m_0$. We reproduce in Fig.~\ref{fig:tb45} and \ref{fig:tb45_muneg} with a double-dashed-dotted magenta line the 95\% C.L. exclusion limit from the 0-lepton and 1-lepton combined analysis reported in Ref.~\cite{atlasbjets}. Notice that the line strictly corresponds to $\tan\beta=40$ and $\mu>0$ -- albeit no substantial variation is expected switching to $\tan\beta=45$ and from the switch in the sign of $\mu$. We show the exclusion limits from the other ATLAS and CMS searches employing the results of Ref.~\cite{nath} that specifically addressed the slice of CMSSM parameter space we consider here, with $\tan\beta=45$.

Fig.~\ref{fig:tb45} shows a summary of the comparison between LHC and dark matter searches for the CMSSM in the large $\tan\beta$ regime, with positive $\mu$. The red regions are excluded by a combination of LEP-II searches for staus and charginos, by the absence of radiative electro-weak symmetry breaking (in the large $m_0$ portion of the plot), and by a stau lightest supersymmetric particle (in the low $m_0$ regions). The dark green shaded region indicates parameter space regions where the thermal relic abundance $0.095<\Omega_\chi h^2<0.13$. The strip to the left of the parameter space, at low $m_0$, corresponds to stau coannihilations; there is then a low-neutralino-mass ``bulk'' region that connects the coannihilation region to the so-called focus point region (see e.g. \cite{focuspoint}), where $\mu\to M_1$ and the lightest neutralino is increasingly higgsino-like. The structure visible in the focus point strip corresponds to the $W$ boson and to the top quark thresholds, at increasing values of $m_{1/2}$.

On this slice of parameter space, the recent results from XENON100 are comparable, and even out-perform, the current LHC reach. Also, large higgsino fractions close to the focus point region make that region accessible, and excluded, by the latest direct detection results, up to neutralino masses of around 200 GeV for models with the correct thermal relic abundance. We take this as a further example of the redundancy and complementarity of LHC and dark matter searches for supersymmetry.

The current neutrino telescope limits from non-observation of high-energy neutrinos from the Sun (a combination of SuperKamiokande and Amanda plus IceCube data, see Ref.~\cite{icecube}) are sensitive only to parameter space points in the focus point region. While they exclude some models with the correct relic density with relatively low masses close to the weak gauge boson masses, typically the required higgsino component to produce enough high-energy neutrinos is too large to have a large enough relic neutralino density. We also show with the orange dashed line the contours corresponding to 180 live-days of data taken with 80 IceCube strings plus DeepCore \cite{icecube} -- a sensitivity result which should likely be achievable on a short time-scale. The future neutrino telescope sensitivity nicely complements the current LHC exclusion limits, and will probe most of the focus point region, even beyond the range of parameters shown in the plot.

We finally show the region excluded by Fermi data from the non-observation of dSph in gamma rays \cite{dsph}: the exclusion limit corresponds to the portion of the plot to the bottom-left of the dark blue line. The current gamma-ray sensitivity is thus not carving into CMSSM models with a ``correct'' relic abundance; also the Fermi sensitivity over this particular parameter space slice is worse than the current low-luminosity LHC reach.

Fig.~\ref{fig:tb45_muneg} explores the negative $\mu$ slice of CMSSM parameter space, for the same value of $\tan\beta=45$ (we use the same color-coding for the red and green regions as in the previous figure). In this case, radiative electro-weak symmetry breaking enforces a much lower value for the mass scale associated to the heavy MSSM Higgs sector. As a consequence, the lightest neutralino can pair-annihilate, in the $v\to 0$ limit, via an $s$-channel exchange of the CP-odd Higgs $A$, since the kinematic condition $m_A\simeq 2m_\chi$ occurs over a viable portion of the $(m_0, m_{1/2})$ parameter space. This so-called ``funnel'' is visible in the figure at $m_0\sim500$ GeV. On resonance, the pair annihilation cross section is so large that the relic thermal neutralinos are under-abundant compared to the dark matter density. That region (and a higgsino-dominated portion of the focus point region) feature large enough pair-annihilation cross sections that the corresponding parameter space is excluded by Fermi data. We emphasize that we assume here that all of the dark matter consists of neutralinos, possibly produced non-thermally (see e.g. \cite{yaguna, moroi}) to account for the underabundant thermal component. We therefore do not re-scale the neutralino number density according to the particle's thermal relic density.

The direct detection prospects for the negative $\mu$ case are slightly worse than for the positive case (due to a different higgsino mixing for the same GUT-scale parameters) in the large $m_0$ region, but are significantly better (due to a lower mass scale associated to the heavy Higgs sector) in the low-$m_0$ region. In either region, direct detection exceeds the LHC sensitivity. Finally, current neutrino telescope data rule out a portion of the parameter space comparable to the LHC exclusion limit for small $m_0$, and rule out some focus point region models with the correct relic abundance.

It is worthwhile to point out that Fig.~\ref{fig:tb45_muneg} illustrates well how current collider and dark matter searches for supersymmetry are exceptionally complementary and synergistic: parameter space regions exist where only one dark matter search technique is at present capable of testing the corresponding supersymmetric model; while in the large $\tan\beta$ regime early searches for supersymmetry with the LHC are out-performed by direct detection and by neutrino telescopes searching for high-energy neutrinos from the Sun, we expect the next round of searches with LHC data to possibly turn this hierarchy around even at large $\tan\beta$. Should a discovery be made with the LHC, one could expect dark matter searches to quickly cover the regions of CMSSM parameter space accessible by the LHC and to readily test options for the underlying fundamental particle theory.

\section{Discussion and Conclusions}\label{sec:disc}

The present analysis considered the supersymmetric parameter space chosen by the ATLAS and CMS collaborations to illustrate how the performance of current early searches for supersymmetry with the LHC compare with the present and near-term performance of experiments sensitive to signals from neutralino dark matter. The chosen theoretical framework, the so-called CMSSM, is a simple albeit somehow representative and motivated choice for how weak-scale supersymmetry might manifest itself; in addition, the CMSSM has been since long the target of, and the theoretical laboratory for, a large body of scientific literature addressing the phenomenology of supersymmetric models from a variety of standpoints.

We specifically considered the performance of direct detection, neutrino telescopes and gamma-ray telescopes in searching for neutralino dark matter; we chose not to consider other indirect detection channels, most notably cosmic-ray antimatter, despite possible intriguing signals \cite{pamela}, due to significant uncertainties on modeling Galactic propagation, that would have made it difficult to reliably assess experimental constraints and capabilities on the theory parameter space. The recent successful deployment of the AMS-02 experiment on the International Space Station will however make antimatter searches for neutralino dark matter extremely compelling in the near future, and will also help reduce the mentioned modeling uncertainties.

Early LHC results out-perform all present efforts searching for neutralino dark matter on the the CMSSM parameter space at low values of $\tan\beta$. Indirect channels are a few orders of magnitude away from current collider capabilities. We illustrated the technical reasons for this hierarchy of sensitivity. We also pointed out that at low $\tan\beta$ the two-photon annihilation mode can be much larger than the naive expectation, even exceeding a branching ratio of 10\% in some CMSSM parameter space regions.

At larger values of $\tan\beta\gtrsim 5$ -- which are both rather natural and phenomenologically not in tension with e.g. LEP searches for the Higgs -- neutralino dark matter detection becomes more competitive. We can here argue that the performance of direct detection experiments (especially with the recent results from XENON100 \cite{xenon100}) is at the same level, on the CMSSM parameter space, as the CMS and ATLAS 35 ${\rm pb}^{-1}$ luminosity sensitivity for values of $\tan\beta\sim20$. For larger values of $\tan\beta$, we found that the CMSSM parameter space covered by direct detection always contains what is currently ruled out by the LHC.

Neutrino telescopes looking for an excess of high-energy neutrinos from the Sun are a particularly promising neutralino search technique for lightest neutralinos with a large higgsino fraction. These, in the CMSSM, correspond to large values of the universal soft-supersymmetry breaking scale $m_0$, a region that is particularly hard to be probed with the collider program. Prospects for the DeepCore plus IceCube80 configuration look particularly bright at large $\tan\beta$, even outperforming spin-independent direct neutralino dark matter searches. The regions of parameter space probed by IceCube are thus highly complementary to the collider program, and partly redundant with direct searches.

Current gamma-ray searches for neutralino dark matter fail to reach levels of sensitivity comparable to what predicted on parameter space regions where neutralino dark matter is produced thermally in the CMSSM. However, if dark matter is mostly consisting of neutralinos, gamma-ray searches probe portions of the CMSSM parameter space that are beyond the reach of the collider program, as well of direct detection and neutrino telescopes. This, again, illustrates how complementary the various ongoing efforts to search for a signature of supersymmetry are.

This study allows us to argue that this is a special point in time, when a remarkable convergence of experimental capabilities (as compared among each other on the simple and practical parameter space of the CMSSM) is achieved in the search for supersymmetry. Should we be on the verge of discovery, this complementarity and redundancy of experimental information will be of essence to constrain the nature of the physical framework hosting the dark matter particle, or to pinpoint the form of low-energy supersymmetry realized in nature.

\begin{acknowledgments}
\noindent  I am thankful to Jason Nielsen for discussions and feedback on this manuscript. This work is supported in part by an Outstanding Junior Investigator Award from the US Department of Energy and by Contract DE-FG02-04ER41268.
\end{acknowledgments}

\end{document}